\documentclass[aps,amssymb,amsmath,amsfonts,twocolumn,showpacs]{revtex4}

\usepackage{graphicx}
\usepackage{epstopdf}
\usepackage{amsthm}
\usepackage{amsmath}
\usepackage{color}

\usepackage{hyperref}
\usepackage{graphicx}
\usepackage{graphics}

\usepackage[T1]{fontenc}
\usepackage[english]{babel}
\usepackage[latin2]{inputenc}

\newtheorem{proposition}{Proposition}
\newtheorem{lemma}{Lemma}
\newtheorem{remark}{Remark}
\newtheorem{theorem}{Theorem}

\newtheorem{corollary}[proposition]{Corollary}

\renewcommand{\>}{\rangle}

\newcommand{\scalar}[2]{\langle #1 | #2 \rangle}

\newcommand{\ket}[1]{| #1 \rangle}
\newcommand{\bra}[1]{\langle #1 |}

\begin{document}
\title{Majorization entropic uncertainty relations}

\author{Zbigniew~Pucha{\l}a$^{1,2}$, {\L}ukasz Rudnicki$^{3,4}$, Karol {\.Z}yczkowski$^{2,3}$}

\affiliation{
\mbox{$^1$Institute of Theoretical and Applied Informatics, Polish Academy of Sciences, Ba{\l}tycka 5, 44-100 Gliwice, Poland}\\
\mbox{$^2$Institute of Physics, Jagiellonian University, ul.\ Reymonta 4, 30-059 Krak\'ow, Poland}\\
\mbox{$^3$Center for Theoretical Physics, Polish Academy of Sciences, al.\ Lotnik\'ow 32/46, 02-668 Warszawa, Poland}\\
\mbox{$^4$Freiburg Insitute for Advanced Studies, Albert-Ludwigs University of Freiburg, Albertstr. 19, 79104 Freiburg, Germany}\\
}


\begin{abstract}
Entropic uncertainty relations in a finite
dimensional Hilbert space are investigated.
Making use of the majorization technique
we derive explicit lower bounds for the sum of R\'enyi entropies 
describing probability distributions associated 
with a given pure state expanded in eigenbases of two observables. 
Obtained bounds are expressed in terms of the largest singular values 
of submatrices of the unitary rotation matrix.
Numerical simulations show that for a generic unitary matrix of size $N=5$
our bound is stronger than the well known result of Maassen and Uffink (MU) 
with a probability larger than 98\%.
We also show that the bounds investigated are invariant under the dephasing and permutation operations. Finally, we derive a classical analogue of the MU uncertainty relation, which is formulated for stochastic transition matrices.
\end{abstract}

\maketitle

{\sl Dedicated to Iwo Bia{\l}ynicki--Birula
on the occasion of his 80th birthday}

\section{Introduction}

The uncertainty principle is often considered as a key feature of quantum theory,
as it explicitly emphasizes the difference with respect to 
its classical counterpart. The original formulation given by Heisenberg in  1927  \cite{He27} which had been devoted to canonically conjugated variables
was further generalized by Robertson in 1929 for arbitrary two observables \cite{Ro29}.
If both observables do not commute, 
it is impossible to specify their precise values  simultaneously.
In this set-up uncertainties  are characterized by 
the variances of both variables, and the relation provides a 
lower bound for the product of these quantities.

Another method to describe the uncertainty
is to use the continuous
entropy of the probability distribution of the measurement outcomes. 
In 1975  Bia{\l}ynicki--Birula and Mycielski derived 
the entropic formulation of the uncertainty relation \cite{BBM75},
in which the main role is played by the lower bound for the sum of two continuous Shannon entropies calculated for position and momentum probability distributions.

Entropic uncertainty relations,  originally introduced for the
infinite dimensional Hilbert space, were later
investigated in the case of a finite dimensional quantum systems.
Consider a pure state $|\psi\rangle$ belonging to an $N$ dimensional 
Hilbert space ${\cal H}_N$ and a
non-degenerate observable $A$, the eigenstates $|a_i\rangle$ 
of which determine an orthonormal basis in ${\cal H}_N$.  
The probability that this observable measured in $|\psi\rangle$ 
gives the $i$--th outcome is $p_i=|\<a_i|\psi\>|^2$. 
The non-negative numbers $p_i$ sum up to unity, $\sum_{i=1}^N p_i=1$,  
so that the properties of the discrete probability distribution $\left\{ p_i\right\}$ can be 
described by the  Shannon entropy $H(p)=-\sum_i p_i\ln p_i$. 

Let $H(q)$ denote the Shannon entropy corresponding to the probability vector
$q_j=|\<b_j|\psi\>|^2$ associated with the second observable $B$. If both observables do
not commute the sum of both entropies for any state $|\psi\rangle$ 
is bounded from below, 
and the bound depends only on the unitary rotation matrix $U_{ij}=\langle a_i|b_j\rangle$. 
The first lower bound:
\begin{equation}
\label{deutsch}
H(p) + H(q) \geq - 2 \ln \frac{1+c}{2}\equiv B_D,
\end{equation}
where $c= \max_{ij} |U_{ij}|$ was with the help of variational calculus derived by Deutsch 
in 1983 \cite{De83}. Maassen and Uffink obtained in~1988~ a stronger result of the form \cite{MU88}
\begin{equation}
\label{MU}
H(p) + H(q) \geq - \ln c^2\equiv B_{MU}.
\end{equation}
Note that for a Fourier matrix of size $N$,
which describes the transition from position to
momentum representation, one has $c=1 / \sqrt{N}$,
so that $B_{MU}=\ln N$.

The result of Maassen and Uffink, while stronger than the
Deutsch lower bound, is known to be not optimal in the general case.
The optimal bound is known only for $N=2$ \cite{SR98,GMR03},
however in higher dimensions the problem remains open.
Let us mention that the entropic uncertainty relations were recently formulated
in various set-ups \cite{BB06,TR11,CYGG11,CCYZ12}. 
To learn more about further developments in that area the reader is
asked to consult reviews  \cite{WW10,BR11}.

In general, it is convenient \cite{BB06}
to work with the R\'enyi entropy
\begin{equation}
\label{eqn:renyi-entropies-probability}
H_{\alpha}(x) = \frac{1}{1-\alpha} \ln \sum_{i=1}^N x_i^\alpha
\end{equation}
which tends to the Shannon entropy for $\alpha \to 1$,
is equal to the min-entropy $-\ln x_{max}$ in the limit $\alpha \to \infty$
and is a non-increasing function of the parameter $\alpha$ \cite{Re61}.
One may then look for bounds for the sum of two  R\'enyi entropies
of  order $\alpha$. 
For instance, explicit bounds in the case $\alpha=1/2$ 
have been for $N=2$ recently obtained by Rastegin \cite{Ra12}.

The aim of this work is to derive a novel bound, 
which for a generic unitary $U$
is with high probability stronger than (\ref{MU}).
We shall establish lower bounds for the sum of two entropies 
of an arbitrary order $\alpha >0$, 
\begin{equation} 
\label{eqn:general-EUR}
H_{\alpha}(p) + H_{\alpha}(q) \geq B_{\alpha}(U),
\end{equation}
with $B_{\alpha}(U)$ depending in general on the whole matrix $U$.

To improve the approaches of Deutsch and Maassen--Uffink (corresponding to the case $\alpha=1$)
we are going to characterize the unitary rotation matrix $U$
by taking into account all its entries.
Our approach is based on the concept of majorization.

Consider any  two probability vectors $x$ and $y$ of sizes $N$ and $M$,
respectively. Associated vectors of size $\max\{N,M\}$,
with coefficients ordered decreasingly and zeros on additional 
coordinates possibly added to the shorter vector,
will be denoted as  $\tilde{x}$ and $\tilde{y}$.
The vector $x$ is said to be {\sl majorized} by $y$, written $x \prec y$, 
if $\tilde{x},\tilde{y}$ satisfy inequalities for all partial sums~\cite{MO79}
\begin{equation}
\sum_{i=1}^m \tilde{x}_i \ \leq \ \sum_{i=1}^m \tilde{y}_i,
\end{equation}
where $m$ runs from $1$ to $\max\{N,M\}$. Note that for $m=\max\{N,M\}$ the inequality is trivially saturated as both vectors sum up to $1$.

The R\'enyi entropy is a Schur--concave function for any parameter $\alpha \geq 0$, what implies that 
if $x \prec y$, then $H_{\alpha}(x) \geq H_{\alpha}(y)$.
In general, when a given function $F$ is Schur--concave 
and two probability vectors satisfy the majorization relation, $x \prec y$,
one obtains the inequality $F(y) \le F(x)$.

This paper is organized as follows. Our main result --- 
the explicit uncertainty relations for the sum of R\'enyi entropies
is derived in section II. In section III we show that
the MU bound and the bounds derived in this work
are invariant with respect to permutation and dephasing operations.
The bounds for some exemplary families of unitary matrices
of size $N=2,3,4,5$ are discussed in section IV. In this section we also use random unitary matrices to compare the precision of various new and previous bounds. Finally, in section V we present a classical analogue
of the Maassen--Uffink relation for an arbitrary
 stochastic transition matrix. 


\section{Main result}

In the entropic uncertainty relation (\ref{eqn:general-EUR}) we bound the sum of entropies of two probability vectors $p$ and $q$. 
However, this sum can be rewritten as the single entropy of the
product vector:
\begin{equation}
\begin{split}
H_{\alpha}(p) + H_{\alpha}(q) &=
 \frac{1}{1-\alpha}\left(\ln \sum_i p_i^\alpha + \ln \sum_j q_j^\alpha\right)\\
&=\frac{1}{1-\alpha} \ln \sum_{ij} (p_i q_j)^\alpha = H_{\alpha}(r),
\end{split}
\end{equation}
where $r = p \otimes q$ is the
tensor product of the classical probability vectors.

Assume that $p$ and $q$ are given by the fixed unitary matrix $U \in \mathcal{U}(N)$ and some
vector $\ket{\psi}$ as: $p_i = |\scalar{i}{\psi}|^2$ and
$q_j = |\bra{j} U \ket{\psi}|^2$, where the vectors $\ket{i}$ for $i=1,\ldots,N$ form the orthonormal basis. It was shown by Deutsch~\cite{De83} that
\begin{equation}
\max_{\ket{\psi},i,j} p_i q_j =
\max_{\ket{\psi},i,j} \left(\frac{p_i+q_j}{2}\right)^2
=\left(\frac{1+c}{2}\right)^2\equiv R_1, \label{EqR}
\end{equation}
where as before $c= \max_{ij} |U_{ij}|$. The above result immediately implies that 
\begin{equation}
r  \prec \left(R_1,1-R_1\right).
\end{equation}
Since the R\'enyi entropies are Schur-concave  we arrive at a first, simple bound
\begin{equation}
H_\alpha(p) + H_\alpha(q) =  H_\alpha(r) \geq \frac{1}{1-\alpha} \ln \left[R_1^{\alpha}+ \left(1-R_1\right)^{\alpha} \right].
\end{equation}

For any rectangular matrix $X$ one defines its spectral norm, equal to its largest
singular value,
\begin{equation}
\| X \| = \sigma_{\max}(X).
\end{equation}
By definition, singular values of $X$ are equal to square roots of the eigenvalues of the positive matrix $X X^{\dagger}$.

\begin{figure}
\includegraphics[scale=0.272]{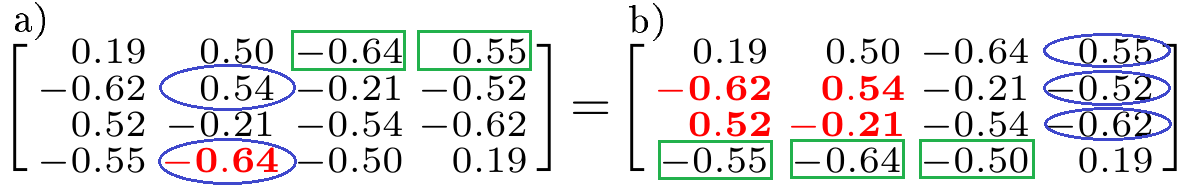}
\caption{Orthogonal matrix $U$ of size $4$ with truncated entries.
To obtain $s_1$ we find its entry with the largest modulus,
denoted in red boldface in panel a). To get $s_2$ we find 
vertical and horizontal $2$-vectors
of the largest norm, marked by blue ellipses and green boxes respectively.
Calculation of $s_3$, shown in panel b) requires comparison of
norms of horizontal (green boxes) and vertical (blue ellipses) $3$-vectors
of $U$ and $2 \times 2$ submatrix (bold entries in red)
with the largest norm.
}
\label{fig:matrix-example}
\end{figure}

Let $\mathcal{A}(m,n)$ denote the set of all $m \times n$ 
submatrices of $U \in \mathcal{U}(N)$,
i.e. the truncations obtained from $U$
by removing arbitrary $N-m$ rows and arbitrary $N-n$ columns. 
Let $A_{m,n}$ denote the {\sl maximal submatrix},
i.e. the element of $\mathcal{A}(m,n)$
with the largest spectral norm.
We shall introduce a set of $N$ coefficients 
\begin{equation}
%
s_k  := \ \max \bigl\{  ||A_{1,k}||,\ ||A_{2,k-1}||, \dots ,
 \  ||A_{k,1}|| \bigr\}
\end{equation}
where the maximum is taken over all submatrices with the same semiperimeter, $m+n=k+1$.
 By construction we have $c = s_1 \leq s_2 \leq \dots \leq s_N=1$, so that $s_1$ 
is equal to the modulus of the largest element of $U$. 
Furthermore, $s_2$ is equal to the maximum of the Euclidean norm of
any two-component part of any column or any row of $U$,
\begin{eqnarray}
s_2 = &&\\
\max && \!\!\!\!\!\!\!\!\left\{ 
	\max_{i,j_1,j_2} \sqrt{ |U_{i j_1}|^2 + |U_{i j_2}|^2 },
	\max_{i_1,i_2,j} \sqrt{ |U_{i_1 j}|^2 + |U_{i_2 j}|^2 } 
\right\}  , \nonumber 
\end{eqnarray}
so it depends only on the moduli of the matrix entries.
In the case of $s_3$ one needs to find the maximum among Euclidean norms of any $3
\times 1$ and $1 \times 3$ vectors and spectral norms of any $2 \times 2$ submatrix 
of $U$ belonging to the set $\mathcal{A}(2,2)$. In the latter case not only
the moduli but also the phases of entries of $U$ become important. In Fig.~\ref{fig:matrix-example} we present an exemplary calculation performed 
for a generic orthogonal matrix of size~4, in which all numbers are truncated up to
two decimal digits.

 In general, one also has a simple bound 
\begin{equation}
s_k \leq  || A_{k,k}||.
\end{equation}
Note that $s_N$ is equal to unity as it is not smaller than the length of any
column (or row) of $U$ and is not larger than the spectral norm of
 the unitary matrix, $\|U\| = 1$.

In the next step we define 
\begin{equation}
R_k = \left(\frac{1 + s_k}{2} \right)^2,
\end{equation}
so that $\left( \frac{1+c}{2}\right)^2 = R_1 \leq R_2 \leq \dots \leq R_N=1$. Let us remind that $R_1$ has been introduced in Eq. (\ref{EqR}).
The above notation allows us to formulate key  results of this paper.

\begin{theorem}\label{th:main-theorem}
For unitary matrix of size $N$ and any normalized vector $\ket{\psi}$ we have
\begin{equation}\label{eqn:main-majorization}
p\otimes q \prec Q,
\end{equation}
where
\begin{equation}
Q = \left(R_1,R_2-R_1,R_3-R_2, \dots, R_N-R_{N-1} \right).
\end{equation}
\end{theorem}

Notice, that from the above theorem we obtain directly the following corollary.
\begin{corollary}
For $Q^{(k)}$ defined as
\begin{equation}
Q^{(k)} = \left(R_1,R_2-R_1,R_3-R_2, \dots, 1-R_{k} \right)
\end{equation}
we have
\begin{equation}
p\otimes q \prec Q = Q^{(N-1)}\prec Q^{(N-2)} \prec \dots \prec Q^{(1)}.
\end{equation}
\end{corollary}

\begin{corollary}
\label{cor:schur}
For any unitary matrix $U$ of size $N$, any normalized vector
$\ket{\psi} \in \mathcal{H}_N$ and a Schur-concave function $F$ we have
\begin{equation}
\begin{split}
F(p\otimes q) &\geq F\left(Q \right)=F\left(Q^{(N-1)} \right) \\
&\geq F\left(Q^{(N-2)} \right) \geq \dots \geq F\left(Q^{(1)} \right).
\end{split}
\end{equation}
\end{corollary}

\begin{corollary}\label{cor:renyi-bounds}
For a unitary matrix $U$ of size $N$, any normalized vector $\ket{\psi} \in \mathcal{H}_N$ and
every $\alpha \geq 0$ we have
\begin{equation}\label{eqn:main-bound}
\begin{split}
&H_\alpha(p) + H_\alpha(q)\geq H_\alpha \left(Q\right),
\end{split}
\end{equation}
what can be extended to
\begin{equation}\label{eqn:K-bounds}
\begin{split}
H_\alpha(p) + H_\alpha(q) \geq B_{\alpha}^{N-1} \geq B_{\alpha}^{N-2} \geq \dots \geq B_{\alpha}^{1},
\end{split}
\end{equation}
with $B_{\alpha}^i = H_\alpha \left(Q^{(i)}\right)$.
\end{corollary}

{\bf Proof of Theorem \ref{th:main-theorem}}:
To prove the majorization relation~(\ref{eqn:main-majorization})
we consider sums of elements of the vector $p \otimes q$,
i.e. $\Xi_k = p_{i_1} q_{j_1} + \dots + p_{i_k} q_{j_k}$ for some indices
$i_1,\dots,i_k$ and $j_1,\dots,j_k$, such that $(i_l,j_l) \neq (i_{l'},j_{l'})$ for $l \neq l'$. 
Assume, that the above sum consists of $m$ different elements of the vector $q$.
If we replace them by the $m$ greatest elements of $q$, i.e.
$\tilde{q}_1,\tilde{q}_2,\dots,\tilde{q}_m$
preserving the order we do not decrease the sum, i.e.
\begin{equation}
\Xi_k \leq \tilde{q}_1 \left(p_{i^1_1}+ \dots +p_{i^1_{k_1}}\right) + \dots + \tilde{q}_m \left(p_{i^m_1}+ \dots +p_{i^m_{k_m}}\right),
\end{equation}
where  $k_1+k_2+\dots+ k_m = k$.  In each parenthesis above we shall next replace components of $p$ by the components of the ordered vector $\tilde{p}$
\begin{equation}
\Xi_k \leq \tilde{q}_1 (\tilde{p}_{1}+ \dots +\tilde{p}_{{k_1}}) + \dots + \tilde{q}_m (\tilde{p}_{1}+ \dots +\tilde{p}_{{k_m}}).
\end{equation}
For all values of the index $i$ we have $k_i \leq k - m + 1$ what provides the final estimate
\begin{equation}
\Xi_k \leq (\tilde{p}_{1}+ \dots +\tilde{p}_{k-m+1})(\tilde{q}_{1}+ \dots +\tilde{q}_{m}).
\end{equation}
The above reasoning gives us the inequality
\begin{equation}
p_{i_1} q_{j_1} + \dots + p_{i_k} q_{j_k} \leq \max_{1 \leq m \leq k} \left(\sum_{l=1}^{k-m+1} \tilde{p}_{l}\right) \left(\sum_{l=1}^{m} \tilde{q}_{l}\right).
\end{equation}

Using the fact, that arithmetic mean is not smaller than the geometric mean we get,
\begin{equation}\label{eqn:AM-GM-ineq}
\left(\sum_{l=1}^{k-m+1} \tilde{p}_{l}\right)\!\!\! \left(\sum_{l=1}^{m} \tilde{q}_{l}\right)
\!\!\leq \frac{1}{4}\left(\sum_{l=1}^{k-m+1} \tilde{p}_{l} +\sum_{l=1}^{m} \tilde{q}_{l}\right)^2.
\end{equation}
Now we can apply Lemma \ref{lemma:main-bound} proven in the Appendix to bound the inner sums by
\begin{equation}
\sum_{l=1}^{k-m+1} \tilde{p}_{l} +\sum_{l=1}^{m} \tilde{q}_{l} \leq 1 + \max_{A \in \mathcal{A}_{k-m+1,m}} \sigma_{\max}(A),
\end{equation}
where for maximum ranges over all sumbatrices of size $(k-m+1) \times m$.
Together with maximization over $m$ we have 
\begin{equation}
\begin{split}
\max_{1 \leq m \leq k} \left( \sum_l^{k-m+1} \tilde{p}_{l} + \sum_l^{m} \tilde{q}_{l} \right) 
\leq 1 + s_k. 
\end{split}
\end{equation}
Thus, we finally obtain the following estimate
\begin{equation}
p_{i_1} q_{j_1} + \dots + p_{i_k} q_{j_k} \leq R_k,
\end{equation}
which gives us the desired majorization relation,
\begin{equation}
p\otimes q \prec \left(R_1,R_2-R_1,R_3-R_2, \dots, R_N-R_{N-1} \right).
\end{equation}
In the maximizing case the inequality~(\ref{eqn:AM-GM-ineq}) can be saturated.
This follows from Remark~\ref{remark:2} in the Appendix, as an optimal choice
of the vector $\ket{\psi}$ implies that both terms in the right hand side
of~(\ref{eqn:AM-GM-ineq}) are equal, so the geometric and arithmetic means coincide.  
\hfill $\Box$

\section{Equivalent unitary matrices and entropic uncertainty}

We say that two unitary matrices $U$ and $V$ are equivalent $U \sim V$ if there exist permutation
matrices $P_1,P_2$ and diagonal unitary matrices $D_1,D_2$, such that
\begin{equation}
V = P_1 D_1 U D_2 P_2.
\end{equation}
It is easy to realize, that any unitary matrix is equivalent to one with real elements
in the first row and the first column which is often called dephased~\cite{TZ06}.
Another simple fact is that any $2 \times 2$ unitary matrix is equivalent to a
real rotation matrix
\begin{equation}\label{eqn:rotation-matrix}
\mathcal{U}(2) \ni U \sim O(\theta) =
\left(
\begin{smallmatrix} 
\cos \theta & -\sin \theta\\
\sin \theta & \cos \theta 
\end{smallmatrix} 
\right) \in SO(2).
\end{equation}

The probability distribution induced by a normalized vector $\ket{\psi}$ is
invariant with respect to diagonal unitary operations. The permutation
matrix changes only the order of coordinates, thus the above equivalence
does not affect the left hand side of relation~(\ref{eqn:general-EUR}).
This observation suggests that one can restrict attention to functions $B(U)$ which are
invariant with respect to the equivalence relation.
 Note that the functions $B_{D}(U)$ and $B_{MU}(U)$, as well as the bounds 
(\ref{eqn:K-bounds}) are invariant with respect to the relation introduced. Therefore, to
analyze the entropic uncertainty relations for unitary matrices of order $N$, 
one can investigate the $N^2-2N-1$ dimensional set of the dephased matrices.

\section{Low dimensional examples}

\begin{figure}
\includegraphics[scale=0.5]{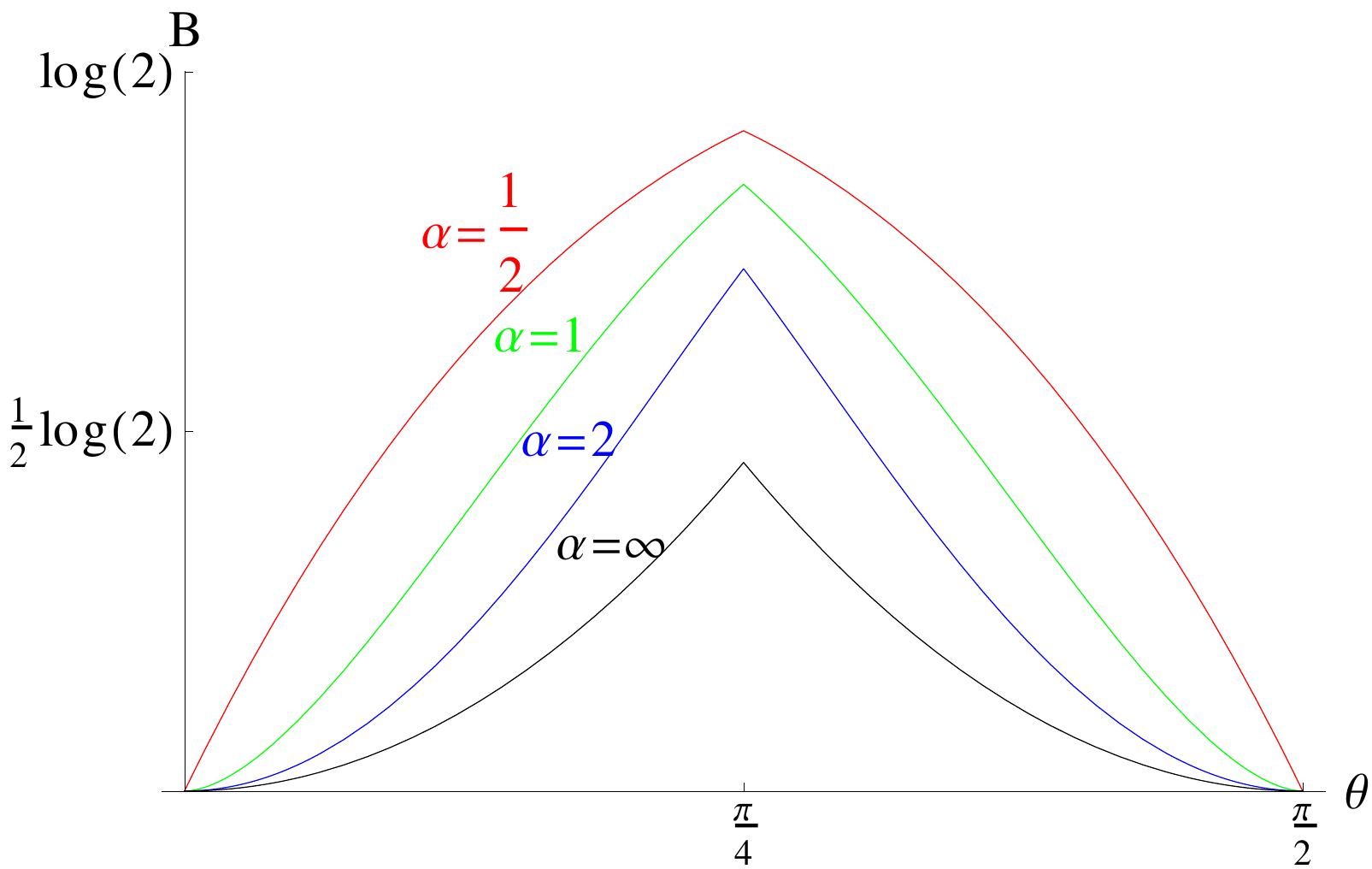} 
\caption{Bound (\ref{eqn:main-bound}) for rotation matrix of size $N=2$, obtained for
different R\'enyi parameters $\alpha=1/2,1,2,\infty$ as a function of
parameter $\theta$.
}
\label{fig:n=2}
\end{figure}

To demonstrate in action the new uncertainty relation proven above consider first 
the case $N=2$. As stated in the previous section it is enough to consider
one-parameter family of rotation matrices since any $2 \times 2$ unitary matrix
is similar to a rotation matrix $O(\theta)$ given in (\ref{eqn:rotation-matrix}).
In Fig.~\ref{fig:n=2} we present the bound (\ref{eqn:main-bound}) for
different values of the parameter $\alpha$.

\begin{figure}
\includegraphics[scale=0.5]{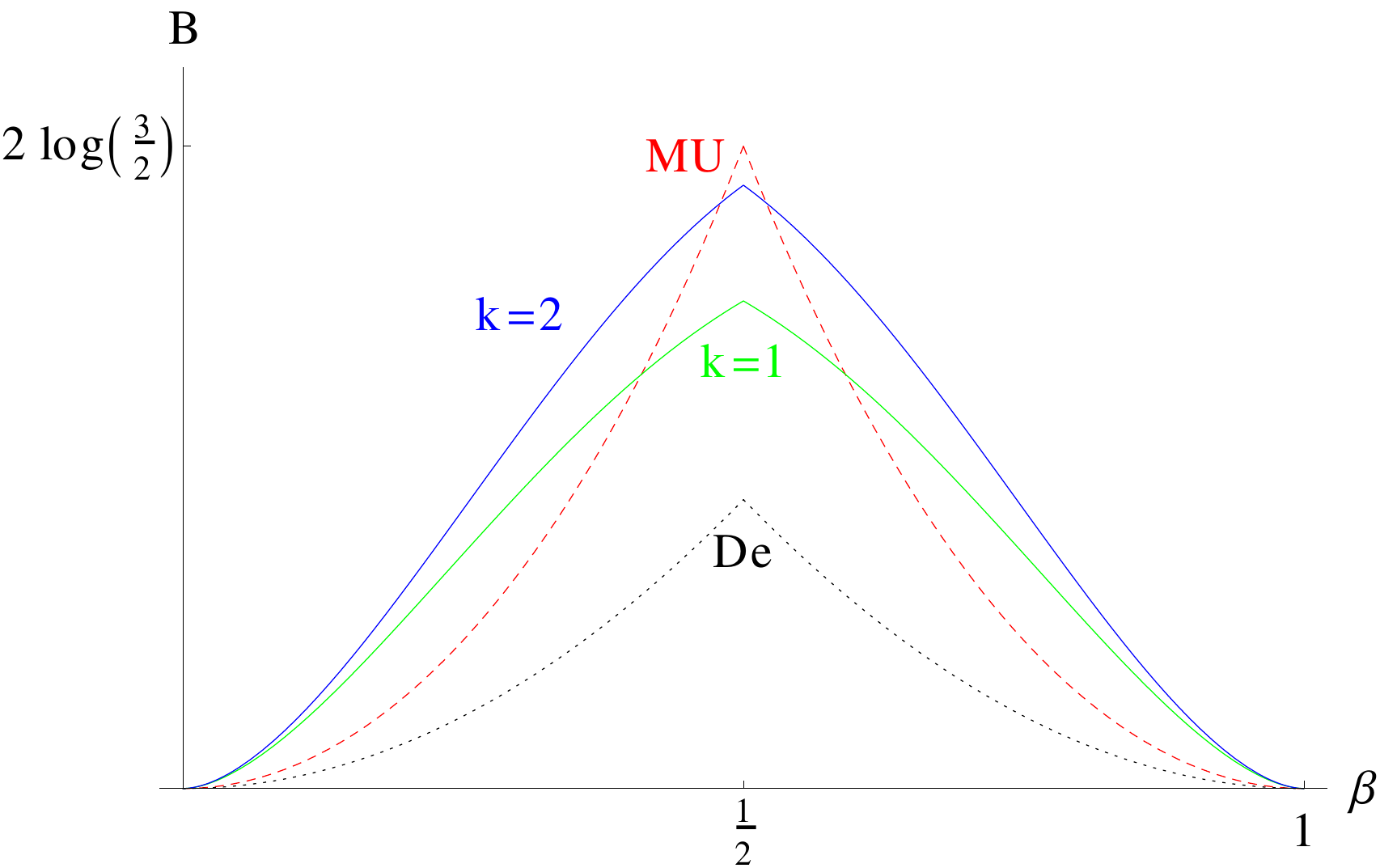}
\caption{ Comparison of bounds (\ref{eqn:K-bounds}) for the family of
matrices $P_3^\beta$ of order $N=3$, where $P_3$ is a circular shift permutation.
Black dotted line represents the Deutsch bound $B_D$~(\ref{deutsch}),
red dashed Maassen--Uffink bound $B_{MU}$~(\ref{MU}) and solid lines represents
bounds $B_1^k$~(\ref{eqn:K-bounds}) for $k=1,2$.}
\label{fig:n=3}
\end{figure}

\begin{figure}
\includegraphics[scale=0.45]{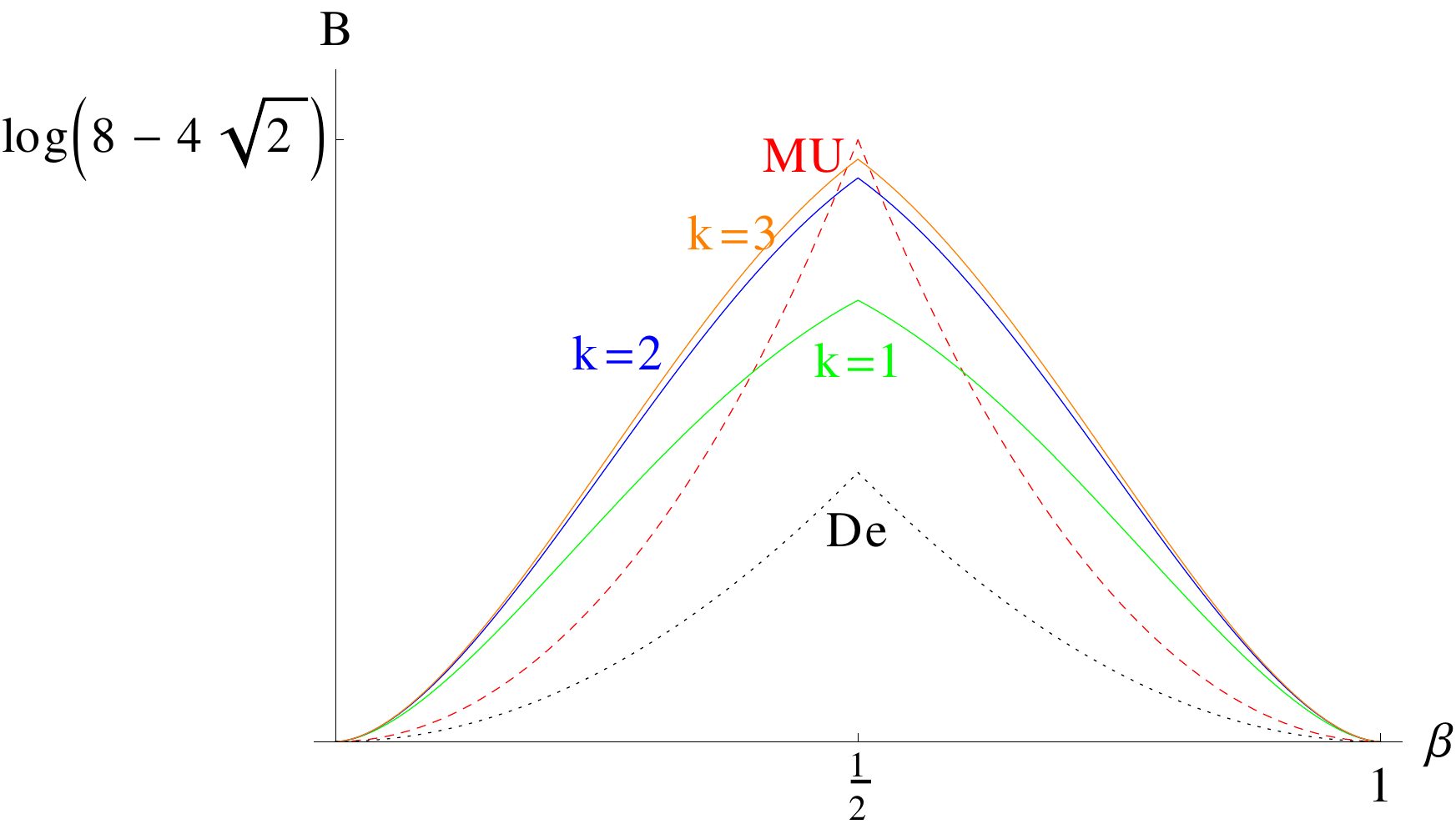} 
\caption{ Comparison of bounds $B_1^k$ for family of matrices $P_4^\beta$, where
$P\in \mathcal{U}(4)$ is a circular shift permutation. Black dotted line represents
Deutsch bound $B_D$~(\ref{deutsch}), red dashed Maassen--Uffink bound $B_{MU}$~(\ref{MU}) and solid lines represents
bounds $B_1^k$~(\ref{eqn:K-bounds}) for $k=1,2,3$.}
\label{fig:n=4}
\end{figure}

\begin{figure}
\includegraphics[scale=0.45]{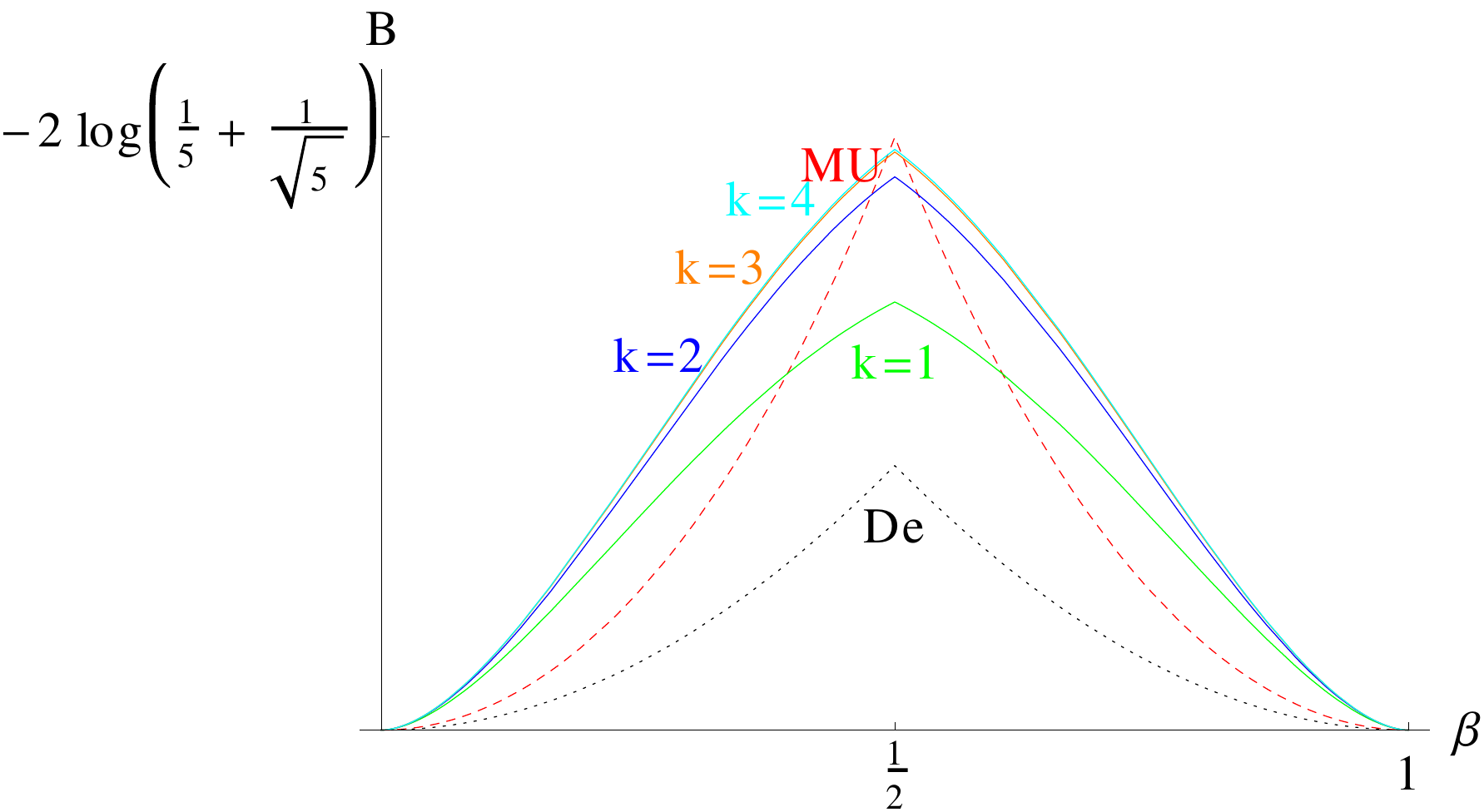}
\caption{As in Fig. (\ref{fig:n=4}) , comparison of bounds $B_1^k$ for family of matrices $P_5^\beta$, where
$P\in \mathcal{U}(5)$ is a circular shift permutation. Black dotted line represents
Deutsch bound $B_D$~(\ref{deutsch}), red dashed Maassen--Uffink bound $B_{MU}$~(\ref{MU}) and solid lines represents
bounds $B_1^k$~(\ref{eqn:K-bounds}) for $k=1,2,3,4$.}
\label{fig:n=5}
\end{figure}

In the case of $N=3,4,5$ we consider a one parameter family of unitary
matrices, given by $P^\beta$, where $\beta \in [0,a]$ and $P$ is a circular shift permutation,
which in the case of $N=3$ reads
\begin{equation}
P_3 =
\left(
\begin{smallmatrix}
0&1&0\\
0&0&1\\
1&0&0
\end{smallmatrix}
\right).
\end{equation}
In Figs.~(\ref{fig:n=3}, \ref{fig:n=4}, \ref{fig:n=5}), we present the
comparison of the bounds~(\ref{eqn:K-bounds}) with the Deutsch bound~(\ref{deutsch}) and
the Maassen--Uffink bound ~(\ref{MU}) for the family of unitary matrices which interpolate
between identity and the $N$-point permutation  matrix $P_N$.

\begin{figure}
\includegraphics[scale=0.8]{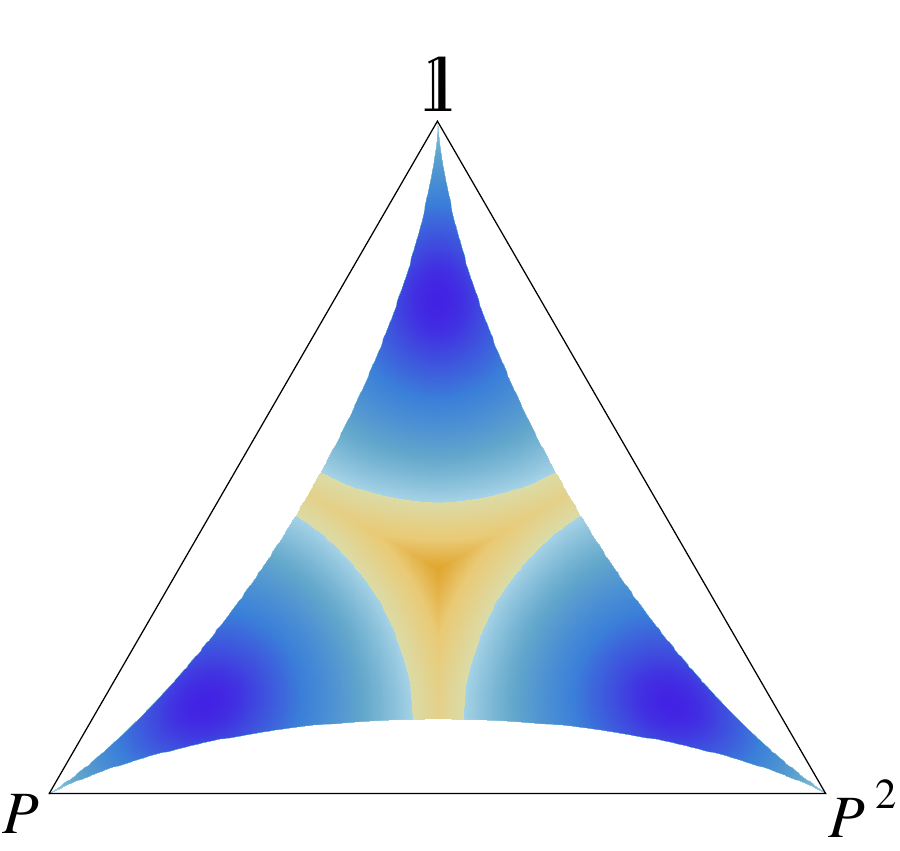}
\caption{ The difference between Maassen--Uffink bound $B_{MU}$~(\ref{MU})
and bound $B_1^2$~(\ref{eqn:main-bound})
for unitary matrices $U$ of size $N=3$ corresponding to the cross-section 
of the set of bistochastic matrices.
In the blue region the MU
bound $B_{MU}$ is lower than the bound $B_1^{2}$, 
while the opposite is true in the yellow region. 
}
\label{fig:unistochastic-simplex}
\end{figure}

In the case $N=3$ we analyze a two dimensional cross-section of the Birkhoff polytope of bistochastic matrices,
and select $\mathcal{B}(a,b) = a P_3 + b P_3^2 + (1-a-b) \mathbb{I}$,
for $0\leq a,b \leq 1, a+b\leq 1$.
Out of this equilateral triangle of bistochastic matrices of order $N=3$ formed by the convex hull of permutation matrices $P_3, P_3^2$ and $ P_3^3={\mathbb{I}}$,
only a proper subset corresponds to unistochastic matrices, such that
there exists a unitary $U$ and $\mathcal{B}_{ij}=|U_{ij}|^2$.
For unitary matrices associated with this subset,
forming an interior of the $3$-hypocycloid \cite{DZ09},
we checked whether the MU bound $B_{MU}$ is larger than the bound $B_1^2$.
Such a set, denoted in light colour in Fig.~\ref{fig:unistochastic-simplex},
contains the center of the figure - the flat bistochastic matrix, $\mathcal{B}_{ij}=1/3$
associated with the Fourier matrix $F_3$ for which the MU bound is sharp.

In order to compare precision of the bound (\ref{eqn:main-bound}) for the
standard case of Shannon entropy $\alpha = 1$ we computed it for random unitary
matrices distributed with Haar measure on $\mathcal{U}(N)$. Probability that for a given
unitary matrix $U$ our bound is better than the Maassen--Uffink bound, increases
with $N$ and reads,
$\mathbb{P}=0.814$ for $N=2$ and $\mathbb{P}=0.971, 0.972, 0.984, 0.991$, for
$N=3,4,5,6$ respectively.  These numbers are obtained numerically
by averaging over samples of $10^7$ random unitary matrices.

\section{Classical analogues of MU bounds}
In the classical case we discuss an $N$-point probability vector $P$ and its
image with respect to a stochastic matrix, $P' = TP$. Stochasticity conditions,
$T_{ij} \geq 0$ and $\sum_i T_{ij}=1$ assure, that $P'$ is also a normalized
probability vector. For a stochastic matrix $T$ and a probability vector $P$,
S{\l}omczy{\'n}ski established~\cite{Sl02} the following inequality for the
Shannon entropy $H$: 
\begin{equation} \label{eqn:clas-ineq-slomczynski}
H^{(p)}(T) \leq H(TP) \leq H^{(p)}(T)+ H(P).
\end{equation}
Here $H^{(p)}(T)$ denotes a statistical mixture of columns of
the matrix $T$ with weights $p_i$, i.e. $H^{(p)}(T) = \sum_i p_i H(\vec{t}_i)$.
Using inequality \eqref{eqn:clas-ineq-slomczynski} we obtain 
\begin{equation}
H(TP) \geq H^{(p)}(T) =  \sum_i p_i H(\vec{t}_i) 
\geq \min_i H(\vec{t}_i),
\end{equation}
which gives us 
\begin{equation} \label{eqn:clasical-bound-H1-H1}
H(P) + H(P') \geq H(P') \geq \min_i H(\vec{t}_i).
\end{equation}
We can continue the estimate and write 
\begin{equation} \label{eqn:clasical-bound-H1-Hinf}
\begin{split}
H(P) &+ H(P') \geq H(P') 
\geq \min_i H_{\infty}(\vec{t}_i) \\
&= \min_i (-\log(\max_j T_{ji})) 
= - \log \kappa,
\end{split}
\end{equation}
where $\kappa = \max_{ij} T_{ji}$.

In this way we obtain an analogue of the Maassen--Uffink uncertainty relation
for the classical maps represented by stochastic matrices. 
The sum of the Shannon entropy of any vector $P$ and the entropy
of its image $P' = T P$ is bounded from below by the logarithm 
of the inverse of the largest element of the transformation matrix.

\section{Concluding remarks}

The problem of establishing optimal entropic uncertainty relations 
for any two observables, the eigenbases of which 
are related by a unitary rotation matrix $U$ of size $N$,
remains open for $N\ge 3$. 

In a recent work of Grudka et al. \cite{G++12}, 
the authors analyzed column (or row) $v$ of $U$, for which
the entropy of the probability vector is the largest;  
$v_i=|U_{ij}|^2$ and $v_i'=|U_{ji}|^2$, where $j=1,\dots N$
and the maximum is taken over $i$. Observe that in this notation
the MU bound (\ref{MU})  reads
$H_1(p) + H_1(q) \ge \max[H_{\beta}(v), H_{\beta}(v')]$ with $\beta= \infty$,
so decreasing the R\'enyi parameter $\beta$
would make the bound stronger.
Unfortunately, numerical simulations show that
an appealing conjecture that the sum of the
Shannon entropies is larger than $H_2(v)$ occurs 
to be true for $N=2$ and $N=3$ only.

On the other hand, in this work we produced a family of
inequalities for the sum of the R\'enyi entropies $H_{\alpha}$
of an arbitrary order which typically are stronger than the
bounds existing in the literature. 
For instance, in the standard case of  $\alpha=1$, 
corresponding to the Shannon entropy,
our result (25) applied to a random unitary matrix
of size $N=3$ gives a bound  stronger  than the  Maassen-Uffink result (2)
for a vast majority of 97\% cases.

It is worth to emphasize that the majorization techniques applied here
enable one to obtain explicit bounds for any Schur--concave 
functions of the probability vector. 
The explicit formulae for the components of the majorizing vector $Q$
derived in this work are expressed in terms of spectral norms of
maximal submatrices of the unitary matrix $U$ analyzed. As this norm is equal 
to the largest singular value of the submatrix~\cite{hj2}, 
our bounds are directly computable.
These bounds are shown to be invariant for any unitary matrices equivalent up  
to permutation and dephasing.

We shall mention that majorization techniques are used in the description of 
quantum entanglement \cite{GL, Partovi, Huang}. For instance the 
bipartite entanglement criteria by G\"uhne and Lewenstein \cite{GL} rely on
 the lower bound for the sum of two R\'enyi entropies. 
Our results can be immediately incorporated in that framework, providing 
sharpened entanglement criteria.

As a side remark we presented a result
analogous to the Maassen--Uffink bound, but formulated for
a classical map described by a stochastic transition matrix $T$.
The sum of the Shannon entropies of an arbitrary  initial 
probability distribution $P$ and its image $TP$ is bounded from below
by minus logarithm of the largest entry of $T$.

\bigskip

Note added. After this work was completed we learned about a very recent results
of Friedland, Gheorghiu and Gour \cite{FGG13}. These
authors independently use majorization techniques
to establish entropic uncertainty relation analogous to (\ref{eqn:main-bound}),
also valid for any R\'enyi entropies and arbitrary Schur-concave functions. 
These powerful bounds can  be used to characterize generalized quantum measurements
described by an arbitrary number of positive operator valued measures (POVM).

\bigskip
 
It is a great pleasure to thank Iwo Bia{\l}ynicki--Birula
for numerous inspiring discussions on quantum theory
we had during the past years. 
We are thankful to P.~Horodecki, M.~Horodecki and {\L}.~Pankowski for 
stimulating discussions and useful correspondence and to V.~Gheorghiu
for drawing our attention to a misprint in the first version of the paper.
We also thank R.~Horodecki and W.H.~{\.Z}urek for encouraging remarks. 
Financial support by the 
NCN grants  number DEC-2011/02/A/ST2/00305 (K\.Z)
and DEC-2012/04/S/ST6/00400 (ZP), and the grant number IP2011 046871
of the Polish Ministry of Science and Higher Education ({\L}R) are gratefully acknowledged.

\appendix

\section{Useful lemma}
We shall present lemma which is the main ingredient of the proof of
Theorem~\ref{th:main-theorem}.
\begin{lemma} \label{lemma:main-bound}
Let $\ket{1}, \ket{2}, \dots, \ket{m} \in \mathcal{H}^N$ and $\ket{a_1},\ket{a_2} \dots \ket{a_n} \in \mathcal{H}^N$
be two orthonormal sets of vectors, then
\begin{equation}\label{eqn:sv-ineq}
\max_{\ket{\psi} \in \mathcal{H}^N} \left(\sum_{i=1}^m|\scalar{i}{\psi}|^2 + 
\sum_{i=1}^n|\scalar{a_i}{\psi}|^2 \right)
= 1 + \sigma_1(A),
\end{equation}
where $\sigma_1(A)$ is the leading singular value of a rectangular matrix $A =
\{a_{ij}\}_{i=1,j=1}^{n,m}$ for $a_{ij} = \scalar{a_i}{j}$ and the maximization is performed over normalized vectors $\scalar{\psi}{\psi}=1$.
\end{lemma}
\begin{remark}
Note, that in the case $m=1$, the matrix $A$ has the form 
\begin{equation}
A = 
\left(
\begin{smallmatrix}
a_{11} \\ 
a_{21} \\ 
\vdots \\
a_{n1}
\end{smallmatrix} 
\right),
\end{equation}
so its norm is equal to the length of the vector, $\sigma_1(A) = \sqrt{\sum_{i=1}^n |a_{i1}|^2}$.
\end{remark}
\begin{remark}\label{remark:2}
One can construct a vector $\ket{\psi_*}$
which maximizes the left hand side of~(\ref{eqn:sv-ineq}). If we denote
\begin{equation}
\begin{split}
\{\ket{\xi_0}, \ket{\eta_0}\} = \arg\!\max[\mathrm{Re} \scalar{\xi}{\eta} : 
&\ket{\xi} \in lin\{\ket{1},\dots,\ket{m}\},  \\
&\ket{\eta} \in lin\{\ket{a_1},\dots,\ket{a_n}
\}],
\end{split}
\end{equation}
we shall take $\ket{\psi_*}$ as a vector proportional to the sum
$\ket{\xi_0} + \ket{\eta_0}$. For this vector one can show that 
\begin{equation}
\sum_{i=1}^m|\scalar{i}{\psi_*}|^2 = \sum_{i=1}^n|\scalar{a_i}{\psi_*}|^2.
\end{equation}
\end{remark}
By $lin\{\ket{1},\dots,\ket{m}\}$ we denote the linear space spanned by the unit vectors $\ket{1},\dots,\ket{m}$. 

{\bf Proof of lemma}:
We begin by rewriting the left hand side of Eq.~(\ref{eqn:sv-ineq}) in terms of matrix multiplication
\begin{equation}
\begin{split}
&\max_{\ket{\psi}} \left(\sum_{i=1}^m|\scalar{i}{\psi}|^2 + \sum_{i=1}^n|\scalar{a_i}{\psi}|^2 \right)\\
&= \max_{\ket{\psi}} \|C \ket{\psi}\|^2 = \sigma_1^2(C) = \lambda_1(C C^{\dagger}),
\end{split}
\end{equation}
where the matrix $C$ is defined as
\begin{equation}
C = 
\left(
\begin{smallmatrix}
\bra{1} \\ 
\bra{2} \\ 
\vdots \\
\bra{m} \\
\bra{a_1} \\ 
\bra{a_2} \\ 
\vdots \\
\bra{a_n} 
\end{smallmatrix}
\right).
\end{equation}
It is easy to calculate, that
\begin{equation}
C C^{\dagger} = 
\left(
\begin{array}{cc}
\mathbb{I}_m & A^{\dagger} \\ 
A & \mathbb{I}_{n}
\end{array} 
\right).
\end{equation}
Next we derive the formula for eigenvalues of matrix $C C^{\dagger}$
\begin{equation}
\lambda_1(C C^{\dagger}) = 1 + \lambda_1\left(
\begin{array}{cc}
0 & A^{\dagger} \\ 
A & 0
\end{array} 
\right) = 1 + \sigma_1(A) .
\end{equation}
The last equality follows from Jordan's definition of singular values and can be
found e.g. in book~\cite{hj2}.
\hfill $\Box$

\end{document}